# Machine Learning in weakly nonlinear systems: A Case study on Significant wave heights in oceans

Pujan Pokhrel
Canizaro Livingston Gulf States Center for
Environmental Informatics
New Orleans, LA 70148, USA.

## ABSTRACT

This paper proposes a machine learning method based on the Extra Trees (ET) algorithm for forecasting Significant Wave Heights in oceanic waters. To derive multiple features from the CDIP buoys, which make point measurements, we first nowcast various parameters and then forecast them at 30-min intervals. The proposed algorithm has Scatter Index (SI), Bias, Correlation Coefficient, Root Mean Squared Error (RMSE) of 0.130, -0.002, 0.97, and 0.14, respectively, for one day ahead prediction and 0.110, -0.001, 0.98, and 0.122, respectively, for 14-day ahead prediction on the testing dataset. While other state-of-the-art methods can only forecast up to 120 hours ahead, we extend it further to 14 days. This 14-day limit is not the forecasting limit, but it arises due to our experiment's setup.

First, this paper proposes an Extra Trees (ET) based machine learning method for forecasting Significant Wave Heights in oceanic waters. Using high-quality CDIP data derived from Datawell Waverider Buoys that provides accurate information about buoy conditions, we show that high instrument precision allows forecasts into the future with high accuracy.

Secondly, we use hv-block cross-validation to make sure that the machine learning algorithms don't see future data while being trained. This allows the models to be robust, which is demonstrated by the performance on the test dataset.

Our proposed setup includes spectral features, hv-block cross-validation, and stringent QC criteria. The proposed algorithm performs significantly better than the state-of-the-art methods commonly used for significant wave height forecasting for one-day ahead prediction. Moreover, the improved performance of the proposed machine learning method compared to the numerical methods shows that this performance can be extended to even longer periods allowing for early prediction of significant wave heights in oceanic waters.

## 1. INTRODUCTION

There has been a growing interest in understanding the physics behind waves and their dynamics in weakly nonlinear media like hydrodynamics, optics, quantum mechanics, Bose-Einstein condensates, nonlinear solids, and finance. These media sometimes experience large, anomalous waves, often known as rogue waves, which are poorly understood. The main way of studying these waves remains deterministic equations like the nonlinear Schrodinger equation, Navier-Stokes, kDV equation, and so on. These equations take a limited number of variables for forecasting, and it is difficult to extend them to macroscopic systems due to computational and time complexity. In recent years, since point measurements over a long duration of time are available for ocean waves through buoys, several efforts are underway to understand wave dynamics in oceans. These waves show analogy in other wave media, and thus help better understand the factors behind wave formation. While Pokhrel et. al have used statistical machine learning methods to forecast anomalous waves, an understanding of the causal factors behind wave dynamics remains unknown. They perform binary predictions on whether the wave is large or not, which does not give the full picture of a continuous process like waves. Intending to study these anomalous waves, it is important to understand the dynamics of the height distribution of the waves via regression approaches.

This paper focuses on ocean waves' accurate prediction, specifically significant wave heights, which remains one of the most important outstanding classical physics problems [1]. While ocean waves are used to study analogies behind various media, nowcasting, and forecasting of waves is also important for myriad other reasons: optimizing ship routes for efficient shipping, avoiding disasters, aiding the aquaculture industry, safely conducting military and amphibious operations by Navy and Marine Corps teams, etc. The other importance of wave prediction lies in efficient renewable energy generated from renewable energy sources like solar, wind, tidal, wave, etc. Commercialization and deployment of wave-energy technologies will require addressing regulatory matters and laws, as well as overcoming technological challenges like accurate and rapid forecasting of ocean conditions and incorporating relevant forecast data in its predictions.

Somayeh [2] used various variables like wind speed, significant wave height, wave period, pressure, air temperature, water temperature, and dew point to predict wave height for up to 24 hours based on the Random Forest method. Mahjoobi and Etemad-Shahidi [3] have used various data mining approaches to predict oceanic wave heights. They have proposed an alternative method for wave hindcasting based on classification and regression trees. The results of decision trees were compared with those of artificial neural networks. The errors of both models were similar, with a nearly equal amount of error. The authors argued that decision trees, with an acceptable range of error, could be used successfully for the prediction of significant wave heights and are better than neural networks in terms of interpretation since they represent rules.

The earlier machine learning approaches used to forecast significant wave heights only take the information from one/few buoys into account. On the data that has Spatio-temporal dependence, only taking the information from a single buoy does not give better performance, no matter how complex the machine learning method is. Also, note that the methods are performing KFold cross-validation which introduces data leakage and is not a proper solution for Spatio-temporal data. Likewise, the work by Christou and Evans [4] explores other data filtering techniques in ocean waves to study rogue waves. These procedures have been used in other studies and found to be reliable through visual inspection [4-8]. These procedures, however, have not been adopted by the machine learning community working on ocean waves. These methods remove outliers and provide a high-quality dataset for training and validation.

While the conclusions of earlier machine learning models have proved to be important in understanding variables that affect wave formation, the fact that they have been trained using noisy data, and a new robust and accurate dataset is available to analyze various wave parameters prompts us to propose a methodology that overcomes the shortcomings of earlier prediction efforts. In addition to that, we present a novel Spatio-temporal cross-validation approach that has not been used by the earlier efforts. We also present other metrics that have been widely used in the physical sciences community so that the machine learning approaches can be compared fairly with numerical approaches.

The main contributions of this paper are as follows.

1) We propose a novel spatio-temporal cross-validation approach for training machine learning approaches on data with Spatio-temporal dependencies, like the ocean waves data from buoys.

2) We compare the statistical methods with LSTMs and find that the improvement offered using deep learning in this data is less than 5% for more than 20 times increase in training time.

3) We use ADF and KPSS tests to show that as the window for forecasts increases (number of points), the data becomes more stationary, and thus more predictable.

4) We re-introduce the data filtering methods, used widely in numerical studies on ocean waves throughout the literature, for the sake of the machine learning community.

## 2. CALCULATION of VARIOUS WAVE PARAMETERS

### 2.1 Significant Wave Height ($H_{m0}$)

Significant Wave Height ($H_{m0}$) is defined traditionally as the mean crest to trough height (mean wave height) of the highest third of the waves. It is also defined as four times the standard deviation of the surface elevation or four times the square root of the zeroth-order moment of the wave spectrum. All these definitions give similar results and may be used interchangeably.

$$H_{m0} = 4 * \sqrt{m_0} \qquad (1)$$

$$m_0 = \sum_{i=0}^{d} E_i(f_i) * d(f_i) \qquad (2)$$

where $m_0$ is defined as the variance of the wave spectrum, $E_i(f)$ is the spectral density at each frequency band, $f_d$ is the dominant frequency at dominant frequency band $d$, $f_0$ refers to the lowest frequency measured by the buoys, which is 0.025 Hz, and $d(f_i)$ is the bandwidth of each frequency band.

### 2.2 Dominant Wave Period ($T_p$)

The dominant wave period is defined as the period corresponding to the frequency band with the maximum value of spectral density in the nondirectional wave spectrum. It is calculated using the formula

$$D_p = \frac{1}{f_p} \qquad (3)$$

where $f_p$ is the peak frequency.

### 2.3 Mean Wave Period ($T_a$)

The mean crossing wave period is defined as the mean wave periods of the waves encountered during the wave sampling period. It is calculated using

$$D_p = \sqrt{\frac{m_0}{m_2}} \qquad (4)$$

where $m_2$ is defined as

$$m_2 = \sum_{i=0}^{d} (E_i(f_i) * d(f_i) * f_i^2) \qquad (5)$$

### 2.4 Mean Zero-Crossing Wave Period ($T_z$)

The zero-crossing wave period refers to the time between the two wave points when the wave height is zero, and the wave is either going up (up-crossing) or down (down-crossing). Likewise, the mean zero-crossing period is defined as:

$$\bar{T}_z = \frac{1}{N} \sum_{i=0}^{N} t_i \qquad (6)$$

where $t_i$ refers to the zero-crossing periods of all waves in the corresponding surface elevation slices (i.e., zero-crossings determined by linear interpolation).

### 2.5 Power Spectral Density (PSD)

In some spectra, the total 'signal energy' is defined as:

$$E_s = \int_{-\infty}^{+\infty} |y(t)|^2 \, dt \qquad (7)$$

where *y(t)* refers to some quantity measured over time, *t*. For example, if *y(t)* is taken to be wave height, then the units of signal energy are $m^2 s$. Note that the signal energy should be a bounded process, and thus, *y(t)* should be a product of a deterministic process, *i.e.*, non-random and with a clear start and finish criteria.

Assuming *y(t)* as an integrable function and Fourier transform of *y(t)*, $Y(f)$ exists, the transformation into the frequency domain is given by

$$Y(f) = \int_{-\infty}^{\infty} y(t) exp^{-i2\pi/t} dt \qquad (8)$$

Following this definition, the power spectral density or signal spectral density is defined as the square of the modulus of $Y(f)$, i.e.,

$$S_E(f) = |Y(f)|^2 \qquad (9)$$

It results in a similar expression to Equation 7 but in the frequency domain. $S_E(f)$ will provide signal energy density at frequency $f$, and if $y(t)$ is wave height, then its unity will be $m^2s^2$. After calculating the spectral density for each frequency, we take the total power spectral density in the wave spectra.

## 2.6 Other Wave Properties

Other wave properties studied were mean wave direction ($\theta$), directional spreading ($\sigma$), skewness, and kurtosis. To calculate various properties of the distribution, we use the method proposed by Kuik *et al.* [15]. The Kuik *et al.* estimate of the kurtosis and skewness used in this paper is based on the integration over the frequency bands from 0.025 Hz to 0.580 Hz over the bulk Fourier moments $a_1, b_1, a_2, b_2$ weighted by the energy density. Note that Fourier moments refer to the coefficients of sines and cosines calculated from the waves after Fourier transform.

$$\begin{bmatrix} a_1 \\ b_1 \\ a_2 \\ b_2 \end{bmatrix} = \frac{1}{E^b} \int_{0.025}^{0.580} \left( df \, E_r(f) \begin{bmatrix} a_1(f) \\ b_1(f) \\ a_2(f) \\ b_2(f) \end{bmatrix} \right) \qquad (10)$$

where $E^b$ is the variance with

$$E^b = \int_{0.025}^{0.580} df \, E_r(f) \qquad (11)$$

afterward, we calculate

$$\theta = tan^{-1}\left(\frac{b_1}{a_1}\right) \qquad (12)$$

$$m_1 = (a_1^2 + b_1^2)^{1/2} \qquad (13)$$

$$\sigma = [2(1 - m_1)]^{1/2} \qquad (14)$$

$$m_2 = a_2 \cos(2\theta) + b_2 \sin(2\theta) \qquad (15)$$

$$n_2 = b_2 \cos(2\alpha) - a_2 \sin(2\alpha) \qquad (16)$$

$$skewness = \gamma = \frac{-n_2}{[(1-m_1)/2]^{3/2}} \qquad (17)$$

$$kurtosis = \delta = \frac{6 - 8m_1 + 2m_2}{[2(1-m_1)]^2} \qquad (18)$$

In (10), the bulk Fourier moments are derived from calculating the skewness and kurtosis. We take the integration of the Fourier moments multiplied by energy [9] and bandwidth. It is then normalized by dividing it with variance calculated in (11). Afterward, we use (12) to (16) to find different parameters, which are used to calculate skewness and kurtosis. Subsequently, we calculate the skewness and kurtosis in (17) and (18), respectively. Note that (12) and (14) give wave direction and directional spread, respectively.

## 3. EXPERIMENTAL SETUP

### 3.1 Dataset

The dataset used in this study contains the buoy data obtained from the CDIP website [10]. The data is obtained in various formats like time-series, 30-min averaged, and spectral data.

We used 1,049,902 points for the training dataset and 449,958 points for the testing dataset, which accounts for 70% and 30% of the total data, respectively. The data was separated based on temporal ordering. Thus, the data in the testing dataset contains the data that is obtained after the time of the data in the benchmark dataset.

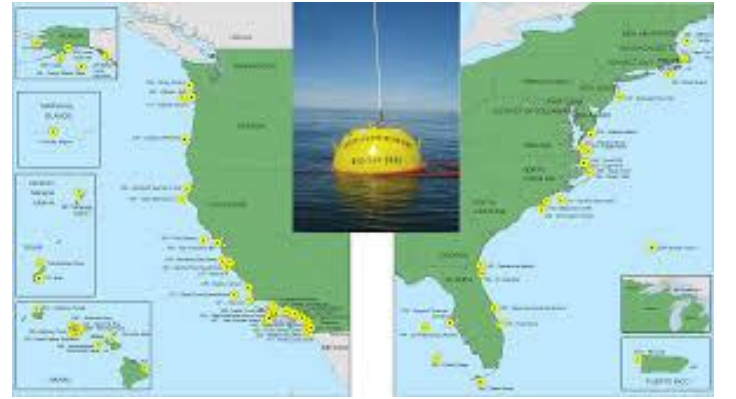

Location of CDIP buoys.

### 3.2 Features

The features used in this study consist of day of the month, time of day, month of the year, significant wave height ($H_{m0}$), mean wave period ($T_a$), dominant wave period ($T_p$), wave direction ($D_p$), Zero up-crossing period ($T_z$), Directional Spread, Power Spectral Density (PSD), depth, skewness, and kurtosis.

**Table 1: Statistical Properties of the dataset**

| Variable | Minimum | Mean | SD | Maximum |
|---|---|---|---|---|
| depth | 11.0 | 287.75 | 522.28 | 1920.0 |
| $H_{m0}$ | 0 | 1.02 | 0.55 | 39.04 |
| $T_z$ | 1.08 | 5.57 | 1.74 | 28.57 |
| $T_p$ | 0 | 10.35 | 4.11 | 40 |
| $D_p$ | -12.07 | 0.04 | 0.21 | 85.1 |
| $T_a$ | 1.82 | 6.35 | 2.10 | 29.18 |
| Spread | 21.76 | 43.26 | 7.14 | 75.36 |
| PSD | 1.16e-5 | 1.80 | 5.33 | 4134 |
| Skew | -770.12 | -1.0e-8 | 1.22 | 587.43 |
| Kurtosis | -966.88 | 1.51 | 4.12 | 502.49 |

### 3.3 Data filtering

Field measurement of waves is subject to various errors that must be removed to obtain a high-quality and reliable dataset. Some examples of the errors include experimental error, buoys getting carried away with the waves, underestimating wave heights, biofouling, etc. Thus, a stringent QC procedure is required to obtain a good dataset in which various machine learning methods can be applied.

The CDIP buoys automatically give the data separated into 30-min intervals. The usage of 30-min intervals lies in how the waves measured are stationary and reasonable predictions can be made. The buoy automatically flags questionable, bad, or CDIP runs missing data points in the same domain as the vertical displacement and other shore-side QC procedures. The 30-min sea with an error flag was removed. For the erroneous data that was not identified by the buoy or CDIP QC procedure, a series of filters were employed to screen and remove the outliers. The filtering procedure has been used widely throughout the literature.

1) Individual waves with zero-crossing wave period > 25.
2) Rate of change of surface elevation Sy exceeded by a factor of 2.

$$Sy = \frac{2\pi\sigma}{Tz}\sqrt{2lnN_z}$$

where $\sigma$ is the standard deviation of the surface elevation $\eta$ and $N_z$ is the number of zero up crossing periods (Tz).

3) Absolute crest or trough elevation greater than 5 times the standard deviation of the 30-min water surface elevation.
4) A single zero-crossing containing > 2304 points.
5) Wave crest elevation $\eta_c > 1.5\ H_s$.
6) Horizontal buoy excursion $\triangle x, \triangle y$ where $\triangle x > 1.8\ H_s$ or $\triangle y > 1.8\ H_s$.

## 3.4 Stationary property of the data

To measure the stationary property of the data, tests like Augmented Dickey-Fuller (ADF) and Kwiatkowski-Phillips-Schmidt-Shin (KPSS) have been proposed.

ADF and KPSS complement each other's strengths in determining if a series is stationary, by proving the existence or absence of unit root. A unit root denotes the stochastic trend (random walk with a drift), which shows a systematic pattern that is not predictable. Note that if the data reject the null hypothesis for ADF, it shows that the series has no unit root. However, in the case of KPSS, rejecting the null hypothesis means that the series has a unit root.

From the dataset, we take

| Num points | ADF | | KPSS | |
|---|---|---|---|---|
| | t-statistic | p-value | t-statistic | p-value |
| 100 | -1.930 | 0.3178 | 1.0950 | 0.0100 |
| 500 | -3.7617 | 0.0033 | 0.1946 | 0.100 |
| 1000 | -5.3625 | 4.0*10^-5 | 0.0901 | 0.100 |
| 5000 | -12.764 | 8.05*10^-24 | 0.0129 | 0.100 |

We similarly perform the same analysis on another buoy to learn the patterns in the data.

| | ADF | | KPSS | |
|---|---|---|---|---|
| Num points | t-statistic | p-value | t-statistic | p-value |
| 100 | -1.03433 | 0.740538 | 1.255329 | 0.01000 |
| 500 | -3.18809 | 0.020684 | 0.14915 | 0.10000 |
| 1000 | -4.865142 | 0.000041 | 0.058471 | 0.10000 |
| 5000 | -13.90092 | 5.74*10^-26 | 0.008604 | 0.10000 |

From Table 2, we can see that although with 100 points, the data rejects the ADF test but does not reject KPSS. However, as the number of points increases, the p-value for ADF of the data points goes down from 0.3178 at 100 points to 8.05*10^-24 at 5000 points.

KPSS stays at the same p-value of 0.0100 even though the number of points increases. This value, however, rejects the null value and shows that the series is stationary.

For the T-statistic of ADF, it decreases from -1.930 at 100 points to -12.764 at 5000 points. Similarly, the T-statistic of KPSS decreases from 1.0950 at 100 points to 0.0129 at 5000 points.

The p-values from Table 2 shows that as the number of points increases, stationary trend can be found through both ADF and KPSS tests. Thus, sufficient data at different times and geographical conditions should allow algorithms to learn the trend such that predictions can be as accurate as possible.

## 3.5 Spatio-temporal stationarity

Definition:

A function f is said to be a stationary-spatiotemporal covariance function on $\mathbb{R}^d * \mathbb{R}$ if it is positive-definite and can be written as:

$$f\big((s;t),(x;r)\big) = C(s-x,t-r), \quad s,x \in \mathbb{R}^d,\ t,r \in \mathbb{R}$$

If a random process $Y(..)$ has a constant expectation and a stationary covariance function $C_\gamma(h;\tau)$, then it is said to be second-order (or weakly) stationary.

Strong stationarity implies the equivalence of two probability measures defining the random process $Y(.)$ and $Y(.+h,.+\tau)$, respectively, for all $h \in \mathbb{R}^d$ and all $\tau \in \mathbb{R}$.

To see if there is any temporal dependence on the data, we compare the covariate-shift on time-delayed outputs.

Covariance Matrix at the buoy at Humboldt bay

| Data | | | | |
|---|---|---|---|---|
| 0-500 | 0.21237 | -0.00445 | -0.00045 | 0.073390 |
| 500-1000 | 0.21238 | -0.17726 | -0.01772 | 0.057513 |

All the coefficients given in Table 4 are similar to each other up to 2 significant digits.

## 3.6 Cross-validation

In KFold cross-validation, the data is first shuffled, then divided into various segments, and machine learning methods are applied thereafter by keeping one part as a testing set and the rest as the training set. In the case of temporal/spatial data, when the machine learning algorithm sees the future data while in training, it learns to remember the outputs and not extrapolate them for predictions. Thus KFold Cross-validation is not suitable for Spatio-temporal data.

### 1) Hv-block cross-validation

The hv-block cross-validation procedure is like the KFold validation, except that there is no random shuffling of the observations. Thus, it renders K blocks of contiguous observations in their natural order. Afterward, while testing the models, adjacent observations between the training and test sets are removed to create a gap between the two sets and increase independence among the observations. The general procedure for hv-block cross-validation is illustrated in Figure 1.

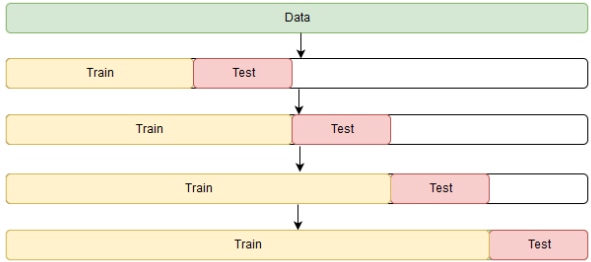

### 2) Spatio-temporal cross-validation

While the hv-block cross-validation can be used with temporal data, in the case of Spatio-temporal data, not considering the spatial components leads to reduced generalization of the machine learning algorithms. The test sets can sometimes only have information from a single buoy, and thus the results can't be generalizable to the whole dataset.

To alleviate this problem, we propose a novel spatio-temporal cross-validation technique. In this procedure, we arrange the data into its spatial and temporal order. The procedure for Spatio-temporal cross-validation is:

1) Sort the data first according to the stations, and then according to the time.
2) Divide the data into K chunks of roughly equal parts.
3) Train the model on the K training sets, in temporal order as shown in Figure 1.
4) Remove some data between the training and test set, and test on the test set as shown in Figure 1.
5) Calculate the error for each fold, and then use it to estimate the averaged error for the whole data.

Note that our proposed cross-validation procedure is similar to hv-block cross-validation. However, since the data is sorted according to both spatial and temporal order before the cross-validation procedure, it captures Spatio-temporal information better than the hv-block cross-validation.

**Figure 1: Setup for hv-blocked cross-validation**

## 3.7 Removing Trend from the Data

To remove trends from the data, we used differencing, a method where instead of predicting the value, the predictor is trained to predict how much the next value will differ from the current one. This step automatically performs standardization of data and removes any trend or seasonality from the data, thus making the predictors more robust and accurate. The procedure of differencing can be shown as

$$\overline{X_1} = X_k - X_0$$
$$\overline{X_2} = X_{k+1} - X_1$$
$$\ldots\ldots$$
$$\overline{X_{n-k}} = X_n - X_k$$

Where $X_n$ refers to the original output values, $\overline{X_n}$ refers to the differenced value for prediction, and $k$ refers to the steps used for predictions. The total data size of $n$ would thus produce $n - k$ points for the training and test dataset.

Afterward, the term that was removed earlier is then added back to the results produced from the machine learning algorithms to produce the final predictions.

## 3.8 Evaluation Metrics

To measure the performance of our model and to compare the results with other methods, we employ various metrics like Root Mean Squared Error (RMSE), Mean Absolute Error (MAE), Variance, R2 Score, Scatter Index (SI), Correlation Coefficient (CC), Bias, and Hanna and Heinold (HH) Indicator. Note that RMSE, Bias and MAE are measured in $meters$, Variance in $meters^2$ and SI, R2 Score, CC, and HH are nondimensional.

**Table 2: Evaluation Metrics and their calculations**

| Name | Mathematical Formula |
|---|---|
| RMSE | $RMSE = \sqrt{\dfrac{\sum_{i=1}^{N}(x_i - \hat{x}_i)^2}{N}}$ |
| MAE | $MAE = \dfrac{1}{N}\sum_{i=1}^{N}|x_i - \hat{x}_i|$ |
| Variance | $Variance = \dfrac{\sum_{i=1}^{N}(x_i - x_m)^2}{N}$ |
| R2 score | $R2 = 1 - \dfrac{\sum_{i=1}^{N}(x_i - \hat{x}_i)^2}{\sum_{i=1}^{N}(x_i - x_m)^2}$ |
| SI | $SI = \dfrac{RMSE}{\frac{1}{n}\sum_{i=1}^{N}x_i}$ |
| CC | $CC = \dfrac{\sum_{i=1}^{N}(x_i - x_m)(\hat{x}_i - \hat{x}_m)}{\sqrt{\sum_{i=1}^{N}(x_i - x_m)^2 \sum_{i=1}^{N}(\hat{x}_i - \hat{x}_m)^2}}$ |
| Bias | $Bias = \dfrac{1}{N}\sum_{i=1}^{N}(x_i - \hat{x}_i)$ |
| HH | $HH = \sqrt{\dfrac{\sum_{i=1}^{N}(x_i - \hat{x}_i)^2}{\sum_{i=1}^{N}x_i\hat{x}_i}}$ |

In the preceding table, $x_i$ refers to the measured value, $\hat{x}_i$ refers to the predicted value at position $i$, $x_m$ refers to the mean of actual values, $\hat{x}_m$ refers to the mean of predicted values, and *N* refers to the number of elements, respectively.

## 4. RESULTS

The machine learning methods were tested for forecasting significant wave height for various forecast ranges. It is noted that 10-fold hv-blocked cross-validation was used for training, and then the optimum parameters were identified. After calculating the best parameters, we train the data on the training set and predict the data using the test set. Note that each window contains 30-min data, and thus, for day-1 forecasting, the algorithm forecasts 48 steps and so on.

We first compare the CDIP and NOAA data since earlier studies use NOAA data. With temporal analysis for buoy at Humboldt bay, we show that both datasets are same, although data obtained from CDIP is of higher resolution. We then perform point forecasts on the Humboldt buoy using hv-block cross-validation. Afterward, we perform point forecasts on the whole data using hv-block cross-validation. Note that hv-block cross-validation only considers the temporal order of the dataset, and not the Spatio-temporal order. We then use the proposed Spatio-temporal cross-validation approach to perform point forecasts. Finally, we compare our methods with the numerical approaches adopted by other weather agencies.

### 4.1 Comparison between data obtained from NOAA buoys, and CDIP buoys

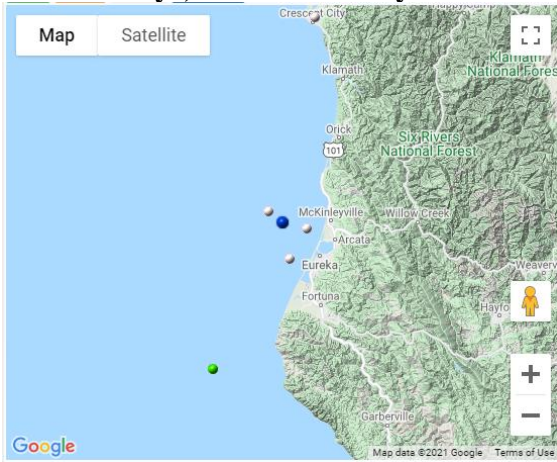

Humboldt bay is represented by the blue dot.

To analyze the discrepancy between the two datasets, and to make sure that they are correct, we compare the data from 02/10/2010 to 02/10/2010 on both buoys and display the value in Table 4.

| Date | Datas. | Hs | Tp | Ta | Dp |
|---|---|---|---|---|---|
| 00:00 | NOAA | 3.01 | 9.88 | 7.46 | 284 |
|  | CDIP | 3.00999 | 9.8765432 | 7.4586081 | 283.59375 |
| 00:30 | NOAA | 3.06 | 10.53 | 7.54 | 288 |
|  | CDIP | 3.05999 | 10.526318 | 7.5419540 | 287.8125 |
| 01:00 | NOAA | 3.29 | 9.88 | 7.65 | 286 |
|  | CDIP | 3.28999 | 9.8765430 | 7.6476054 | 286.20625 |
| 01:30 | NOAA | 3.05 | 9.88 | 7.29 | 292 |
|  | CDIP | 3.04999 | 9.8765430 | 2.2870130 | 292.03125 |

We also scan for other data points during the period (12/10/2017).

| Date | Datas | Hs | Tp | Ta | Dp |
|---|---|---|---|---|---|
| 00:00 | NOAA | 1.59 | 13.33 | 10.12 | 287 |
|  | CDIP | 1.59000 | 13.3333 | 10.12227 | 286.8125 |
| 00:30 | NOAA | 1.40 | 14.29 | 10.01 | 285 |
|  | CDIP | 1.39999 | 14.28571 | 10.00873 | 281.1875 |
| 01:00 | NOAA | 1.41 | 11.76 | 9.75 | 281 |
|  | CDIP | 1.40999 | 11.76471 | 9.748840 | 281.1875 |
| 01:30 | NOAA | 1.52 | 13.33 | 10.25 | 287 |
|  | CDIP | 1.51999 | 13.33333 | 10.249608 | 281.1875 |

While we don't show the comparison between all the data points, it should now be obvious for the reader to see that CDIP data has a higher resolution than the CDIP dataset. We note that CDIP data is given in NetCDF format which uses float64 format to store the floating points. However, since NOAA provides the data in text format, the data is also stored with reduced precision and needs to be clipped to two/three decimal places.

### 4.2 Point forecasts on a single CDIP buoy

First of all, we take the data from a single buoy and perform forecasts. Table 5 shows the performance of ET algorithm for short-term prediction (up to 24 hours).

| Time Step (hours) | Differenced | |
|---|---|---|
|  | RMSE | $R^2$ |
| 1 | 0.018 | 0.999 |
| 3 | 0.291 | 0.911 |
| 6 | 0.423 | 0.822 |
| 9 | 0.490 | 0.759 |
| 12 | 0.567 | 0.695 |
| 24 | 0.686 | 0.552 |

Table 4 shows the performance of ET algorithm for forecasts up to longer durations.

| Time Step (days) | Differenced | |
|---|---|---|
| | RMSE | $R^2$ |
| 1 | 0.686 | 0.552 |
| 3 | 0.883 | 0.273 |
| 6 | 0.903 | 0.225 |
| 9 | 0.911 | 0.217 |

Since the performance of the ET algorithm is better on the dataset obtained after CDIP's robust cross-validation than the one from NOAA buoys, we use the CDIP dataset for further experiments.

While it is evident that there is no discernible difference between the datasets, and the main difference is the features provided, we perform further experiments with the CDIP data. The use of spectral data from CDIP buoys also allows us to calculate various features that are important for the prediction of significant wave heights.

We then perform short-term predictions on the dataset in 30-minute intervals for up to 24 hours in 3-hour intervals. Note that for every 3 hours ahead prediction, our framework performs 3*2 = 6 prediction because the resolution of time-averaged spectral data is 30 minutes for CDIP buoys.

## 4.3 Tests on the whole data with hv-block cross-validation

We then compare the performance of Gaussian Model Mixture (GMM) as a benchmark to compare the performance of LightGBM and GMM based machine learning models.

The performance of GMM stabilizes around RMSE of 0.67 and R2 of 0. LightGBM similarly, stabilizes around RMSE of 0.399 and R2 of 0.476.

Afterward, to train the dataset on the bigger dataset, we take the data from buoys mentioned in Section 3.1. Note that the performance of the machine learning algorithm increases when tested on the complete data.

| Time Step (days) | GMM | | LightGBM | |
|---|---|---|---|---|
| | RMSE | $R^2$ | RMSE | $R^2$ |
| 1 | 0.485 | 0.434 | 0.324 | 0.657 |
| 3 | 0.634 | 0.037 | 0.388 | 0.505 |
| 6 | 0.660 | -0.02 | 0.395 | 0.488 |
| 9 | 0.675 | -0.076 | 0.399 | 0.479 |
| 12 | 0.667 | -0.054 | 0.396 | 0.480 |
| 24 | 0.671 | -0.074 | 0.400 | 0.476 |

From the result of GMM that stabilizes around 0.67 with $R^2$ the score of 0, and the result of LightGBM that has RMSE of 0.40 with $R^2$ of 0.476 suggests that even a properly calibrated straight line should have RMSE around 0.67. The LightGBM model is only performing better because it has captured some of the Spatio-temporal correlations in the data.

To visualize the data structure properly, we run a Linear Regression model on the whole dataset. The coefficients of various features, and the T-statistic and p-value are shown in Table .

| Feature | Coefficient | T-Statistic | p-value |
|---|---|---|---|
| Latitude | | | |
| Longitude | | | |
| Year | | | |
| Month | | | |
| Day | | | |
| Time | | | |

Since the p-value for depth, time variables, latitude, and longitude are less than 0.01, we conclude that the data has a complex Spatio-temporal structure. We then perform hv-block cross-validation and the proposed Spatio-temporal cross-validation. Note that the hv-block cross-validation only takes the temporal component of the data into account, unlike the proposed method. The proposed method first rearranges the dataset by time and spatial location and then performs cross-validation.

## 4.4 Tests on the whole data with proposed Spatio-temporal cross-validation

| Hours | LightGBM | | ET | |
|---|---|---|---|---|
| | RMSE | $R^2$ | RMSE | $R^2$ |
| 1 | | | | |
| 3 | 0.126 | 0.983 | 0.173 | 0.969 |
| 6 | 0.142 | 0.973 | 0.246 | 0.937 |
| 9 | 0.147 | 0.978 | 0.297 | 0.909 |
| 12 | 0.148 | 0.977 | 0.338 | 0.884 |

We now perform long-term forecasts.

| Days | LightGBM | | ET | |
|---|---|---|---|---|
| | RMSE | $R^2$ | RMSE | $R^2$ |
| 1 | 0.162 | 0.941 | 0.196 | 0.904 |
| 3 | 0.160 | 0.940 | 0.200 | 0.900 |
| 6 | 0.163 | 0.936 | 0.199 | 0.901 |
| 9 | 0.174 | 0.925 | 0.208 | 0.892 |
| 12 | 0.171 | 0.927 | 0.203 | 0.896 |

| 24 | 0.163 | 0.934 | 0.196 | 0.904 |

## 4.5 Comparison of statistical and deep learning methods for 1-day ahead forecasting

| Method | RMSE | $R^2$ |
|---|---|---|
| LightGBM | 0.162 | 0.941 |
| LSTM | 0.150 | 0.943 |

While the LSTM model achieves RMSE of 0.150, the LightGBM model has an RMSE of 0.162. The training time for the LSTM model was 10 hours compared to less than 5 minutes for other methods. Training a deep learning method that takes more than 20 times time for less than 5% improvement in RMSE did not seem like an attractive option regarding the main thesis of the paper, which deals with data filtering and cross-validation approach. In concern of computational complexity and interpretability, we discard the deep learning model and continue with the LightGBM model for final predictions.

Our thesis deals with the novel cross-validation approach and not a novel machine-learning algorithm. Thus, we do not deem it important to compare individual (previous) state-of-the-art approaches tried on ocean waves with the proposed cross-validation approach. These methods have been trained on a handful of NOAA buoys that provide wind, and current information by default. It is certainly possible that their methods are equal/better than ours with more features, but it is not feasible to make planetary-scale forecasts with those few buoys. We then continue with the comparison with numerical approaches for 1-day ahead prediction.

## COMPARISON OF SIGNIFICANT WAVE HEIGHT PREDICTION WITH OTHER STATE-OF-THE-ART METHODS

In this section, we compare our method with other commonly used methods for 1 day ahead forecasting. Note that since Bidlot et *al.* [11] tested the other methods using buoy data from June to August 2007, we utilize data from January 2001 to June 2007 as a training dataset and June to August 2007 as the test dataset. The models compared are European Centre for Medium-range Weather Forecasts (ECMWF) [11-13], Met Office (MO) [14], Fleet Numerical Meteorology and Oceanography Centre (FNMOC), Meteorological Service of Canada (MSC) [15-17], National Centers for Environmental Prediction (NCEP) [18-21], Meteo France (MF) [22-25], Deutscher Wetterdienst (DWD) [26], Bureau of Meteorology (BoM) [11, 27-29], Service Hydrographique et Oceanographique de la Marine (SHOM) [30], Japan Meteorological Agency (JMA) [31], Korea Meteorological Administration (KMA) [32]. Note that all the methods compared are global models. Note that the forecasting is done over 12-hour mean values.

**Table 3: Comparison of the proposed method with other state-of-the-art methods**

| Method | SI | Bias(m) | CC | RMSE(m) |
|---|---|---|---|---|
| ECMWF | 0.151 | -0.02 | 0.95 | 0.25 |
| MetO | 0.210 | 0.20 | 0.92 | 0.40 |
| FNMOC | 0.192 | 0.04 | 0.94 | 0.32 |
| NCEP | 0.186 | 0.11 | 0.94 | 0.33 |
| MF | 0.231 | 0.22 | 0.89 | 0.44 |
| DWD | 0.202 | 0.04 | 0.92 | 0.34 |
| BoM | 0.226 | 0.03 | 0.89 | 0.38 |
| SHOM | 0.180 | 0.00 | 0.93 | 0.30 |
| JMA | 0.209 | -0.18 | 0.91 | 0.40 |
| KMA | 0.305 | 0.05 | 0.79 | 0.50 |
| **Our Method** | **0.130** | **-0.002** | **0.97** | **0.14** |

In the preceding table, the best values are highlighted in bold. Note that our proposed method has a very low Scatter Index (SI), bias, and RMSE value. Likewise, the Correlation Coefficient (CC) of the proposed method is very high compared to the other methods, suggesting that the proposed method can be used to forecast the significant wave height in oceans.

While the proposed method outperforms these state-of-the-art methods for select buoys, it is important to note that the buoys are not always at the points from which the grid is initialized. Thus, numerical simulations would have a lower RMSE if the comparisons are made that way. However, our result shows that the wave systems are weakly nonlinear and second-order stationary, and certain predictions can be made up to a long time into the future, to a reasonable degree of accuracy.

## 5. DISCUSSION

The results section shows that in general, for both Extra Trees and LightGBM, the performance of the algorithms decreases sharply from day 0 to day 1 but remains consistent afterward with small fluctuations. The small fluctuations can be attributed to the feature of the dataset and, to some extent, to the noise inherent in the dataset. The Extra Trees, which performs bagging on the data, performs better than LightGBM, which performs boosting on the data. The comparatively better performance of ET shows that the data contains more variance error than bias, which is often due to noise in the dataset.

Next, we plot the feature importance, computed by the Extra Trees algorithm for significant wave heights. The results obtained are shown in Figure 10.

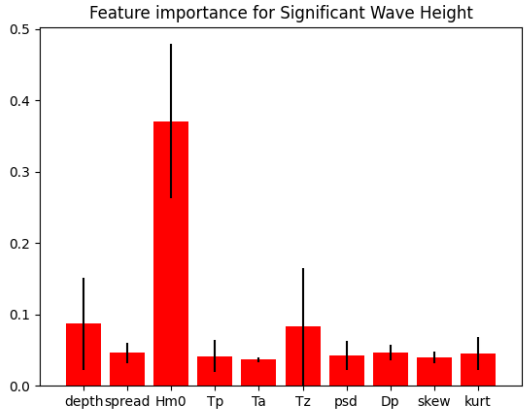

**Figure 10: Feature importance of various features for 1 day ahead forecasting of significant wave heights**

In Figure 10, significant wave height has the highest feature importance of 0.370. Depth has the second-highest feature importance of 0.087. Directional spread, zero-crossing period, dominant wave period, power spectral density, mean wave period, wave direction, skewness, and kurtosis have feature importance of 0.046, 0.042, 0.036, 0.083, 0.042, 0.047, 0.040, and, 0.045 respectively.

The two machine learning methods chosen in this paper are LightGBM, and ET, which minimize bias and variance, respectively. LightGBM performs boosting on the data, and ET performs bagging. The superior prediction capability of ET algorithm suggests that the data is noisy.

First, we use various data filtering methods to get the most accurate data for training and test purposes. These methods are widely used in physical oceanography to filter outliers in the buoy wave data. We use hv-blocked cross-validation to ensure that the data's temporal order is not disturbed while training the model. After that, we compare various machine learning algorithms on the features generated and found that ET, a tree-based machine learning algorithm, works better for forecasting using various wave features derived from the buoys.

Our proposed algorithm has a better predictive capability for forecasting wave height than other state-of-the-art methods, both statistical and numerical. Despite using only, the variables available from the spectra of the buoys, our method performs better than other methods that take wind speed and air/water temperature, and other variables into account. Our algorithm's superior performance can be attributed to cross-validation, data filtering, and differencing.

In this paper, we first introduced various data filtering methods and time-based hv-block cross-validation for use in wave data. Since the instrument precision of CDIP buoys is higher than that of NOAA, the performance of machine learning algorithms is higher. Thus, the data from CDIP buoys seem to be a good choice for future prediction efforts.

We also proved that the data is stationary when taken the number of samples increases with the use of ADF and KPSS tests. Likewise, while testing the hypothesis that whether the data is Spatio-temporal stationary or not, we found that predictions can be made without decreasing accuracy if we take the geography into account and the data from other buoys at various locations.

For 1-day ahead prediction, even though we have better performance than other state-of-the-art machine learning approaches, we don't claim to outperform them. Our proposed methodology involves taking Spatio-temporal data, and robust cross-validation for such data. Without the use of Spatio-temporal data, and proposed cross-validation methodology, our methods also have similar performance as the other state-of-the-art approaches. The performance of our algorithm is better with data from more buoys because we handle Spatio-temporal weakly-nonlinear stationary data better with proposed novel Spatio-temporal cross-validation, and not due to the complexity of the model. Unlike those approaches, we take the data from surrounding buoys which helps improve performance, and our cross-validation methodology allows us to estimate the performance of statistical approaches while avoiding data leakage. Similarly, although our method outperforms the numerical state-of-the-art approaches, it would be dishonest to claim to outperform them. Numerical approaches divide the Earth into grids and then perform calculations over a point in the grid. The calculations are then used to interpolate various wave properties within the grid. Thus, the data from the numerical methods only show the mean value for the grid. This scheme is used because of the computational and time complexity of the numerical methods.

For longer horizon forecasts, however, the performance of the proposed algorithms does not decrease with time. With the use of ADF and KPSS tests, we showed that the data is stationary in time. Similarly, with the use of covariate shift and expectation, we can see that the data is second-order (weakly) stationary. Likewise, with the use of 2D Kernel Density Estimation (KDE), we can see that the data has spatial dependence.

In this paper, we propose a novel spatio-temporal cross-validation technique for use in oceanic data. We first show that the data is stationary using ADF and KPSS tests. While our proposed method outperforms the other numerical approaches, claiming that the method is superior to others would be a dishonest claim. Numerical methods discretize the planet into grids and make predictions from the specified point, which is then interpolated for other points in the grid. Since the buoys are not always at the center of those grids, the performance of numerical methods might differ on various buoys. However, our models were trained on the buoys themselves, and even then, the machine learning algorithms had similar performance to numerical approaches. However, our results suggest that machine learning algorithms should adopt ideas from physics to enforce constraints on functions being optimized for further improvements.

## 6. CONCLUSION

While the performance of the machine learning algorithms seems counter-intuitive, recall that CDIP buoys perform robust shore-side QC procedures, in addition to the automated procedure used by buoys. The training data also contains information from neighboring buoys, and the LightGBM algorithm can capture various nonlinear features. In addition, significant wave height is defined as one-third of the highest waves in oceans for a duration, or equivalently, 4 times the zeroth moment of the spectra. It is itself a statistical definition. Thus, we are not predicting the instantaneous accelerations of the buoys, but the mean of the motion over the period of time in which it is stationary.

In theory, wave forecasting can be done using model equations of empirical relationships between various wave parameters.

Compared to the nonlinear differential equations, machine learning methods provide similar models with lower computational complexity and computation time.

This study analyzed various methods for quality control of data and cross-validation for time data. The methods, combined with the machine learning algorithms, allowed us to forecast significant wave heights for various time ranges without an abrupt decrease in prediction performance over time. Likewise, the use of tree-based algorithms explored in this paper (LightGBM and ET) give feature importance, which helps quantify the impact of the features and helps understand how the algorithm is making decisions. Moreover, the data collected from CDIP buoys are very accurate after the application of the QC procedure, thus allowing for improved performance. The QC methods, block cross-validation, and machine learning methods help us predict significant wave heights, a nonlinear phenomenon with performance comparable to the state-of-the-art methods. Future studies will incorporate ocean currents, wind, and other wave parameters for wave and wind forecasting.

While showing the analogies behind the wave formation in other media remains beyond the scope of the paper, we posit that similar algorithms and setups would allow for forecasting wave heights and other wave properties on various other nonlinear media allowing for early forecasting of such waves.

# 7. ACKNOWLEDGMENTS